# Broad-temperature-range ultrafast terahertz excitation of collective dynamics in polar skyrmions

Wei Li[a], Sixu Wang[a], Pai Peng[b,c], Haojie Han[a], Xinbo Wang[b]*, Jing Ma[a], Jianlin Luo[b], Jun-Ming Liu[d], Jing-Feng Li[a], Ce-Wen Nan[a], Qian Li[a]*

[a] *State Key Laboratory of New Ceramics and Fine Processing, School of Materials Science and Engineering, Tsinghua University, Beijing 100084, China*

[b] *Beijing National Laboratory for Condensed Matter Physics, Institute of Physics, Chinese Academy of Sciences, Beijing 100190, China*

[c] *State Key Laboratory of Low Dimensional Quantum Physics, Department of Physics, Tsinghua University, Beijing 100084, China*

[d] *Laboratory of Solid State Microstructures and Innovation Center of Advanced Microstructures, Nanjing University, 210093, Nanjing, China*

* Corresponding author: (QL) qianli_mse@tsinghua.edu.cn; (XW) xinbowang@iphy.ac.cn

**Abstract**

Ultrafast coherent control of electric dipoles using strong terahertz (THz) pulses provides a means to discover hidden phases of materials and potentially leads to applications in high-speed electro-optic devices. The effectiveness of this means, albeit demonstrated in architype (incipient) ferroelectric systems such as $SrTiO_3$, hinges on a spectral overlapping between their soft phonon modes within the excitation bandwidth of THz pulses. Generally this can only induce an appreciable coupling close to the phase transition temperatures, where the lattice phonons substantially soften. Because of their emergent subterahertz collective dynamics, a fundamentally distinct and effective THz coupling mechanism can be envisaged in topological polar structures recently discovered in $PbTiO_3/SrTiO_3$ superlattices. Here, we show that polar skyrmions can be coherently driven into a hidden phase with transient macroscopic polarization, as probed based on THz field-induced second harmonic generation and optical Kerr effects. Such an ultrafast THz-driven phase transition is found to sustain across a broad temperature range of 4-470 K, in accordance with the equilibrium stability field of the skyrmions. Supplemented by dynamical phase-field simulations, we identify the spatial symmetries and relaxation behaviors of the excited collective modes, thereby revealing their correlation with the emergence of the polar phases. Our results unveil the exotic dynamical properties of topological polar structures, which could be technologically exploited given their remarkable flexibility in structure design and tunability under external fields.

**Main**

In epitaxial ferroelectric/dielectric superlattices, as exemplified by the $(PbTiO_3)_m/(SrTiO_3)_n$ ($m$, $n$ are numbers of the unit-cell layer) system, precise tuning of the layer thickness and substrate mismatch strain has been discovered to stabilize nanoscale topological phases with noncollinear polarization vector configurations and long-range in-plane ordered states[1–6]. These exotic phases, such as polar vortex, skyrmion and dipole wave, mainly reside in the constituent ferroelectric layers and exhibit a series of fascinating phenomena rarely seen in conventional ferroelectric materials, including negative capacitance[7,8], optical chirality[9–12] and light-induced supercrystal phase[13,14]. However, compared to their magnetic counterparts[15], the dynamical behaviors of polar topological phases are yet to be systematically explored to unveil emerging functional properties[16–19]. Microscopy studies have examined the *in-situ* polarization evolution processes of the polar vortices and skyrmions, revealing their transitions into different topologies or topologically trivial phases under the action of external electric, mechanical or thermal fields[20–25]. Ultrafast diffraction studies have revealed that polar vortices host a set of collective vibrational modes within the sub-terahertz (THz) frequency range, akin to the infrared-active optical phonons of the underlying lattices[26]. Such collective modes provide unusual dielectric resonance mechanisms that potentially can be harnessed for electrically tunable devices operating at millimeter-wave and THz frequencies[27–29]. Besides, the enhanced structural stabilities of these phases, a topological protection attribute, can foster thermally stable macroscopic functional properties with broadened application scopes. This could be particularly relevant for polar skyrmions whose polarization configuration is of stronger three-dimensional characters and better topological phase stability, compared with polar vortices[8]. The collective dynamics of the former system has not been addressed thus far.

Ultrafast coherent manipulation of electric dipoles has been demonstrated in architype perovskite ferroelectrics, including $SrTiO_3$, $KTaO_3$ and $BaTiO_3$, using single-cycle, strong-field THz pulses via direct electrostatic coupling effects. In these systems, crucial to an effective coupling is the spectral overlapping between the THz pump pulses and their specific lattice phonon modes thus to realize a resonant excitation state[30–36]. Phenomenologically, the free-energy functional of ferroelectrics can be described with respect to the soft phonon mode coordinate $\boldsymbol{Q}_{FE}$ responsible for the polarization order parameters[37]. The polarization of PbTiO$_3$ arises from the condensation of its Last mode, in which the displacement of Pb$^{2+}$ ions is out-of-phase with those of Ti$^{4+}$ and O$^{2-}$ ions[38]. Below the Curie temperature, the double-well free-energy landscape posits $\boldsymbol{Q}_{FE}$ at a specific value corresponding to the equilibrium polarization[39]. Upon THz excitations, the $\boldsymbol{Q}_{FE}$ shifts periodically as a function of the external electric field and thereby triggers polarization oscillations. Generally, displacive ferroelectrics exhibit strongly temperature-dependent soft mode dynamics and thus the spectral overlapping with THz excitations is confined to the Curie temperature, beyond which the mode frequencies quickly recover[38,40–42]. The long-range dipolar correlation characters also lead to large potential well depths, rendering the switching of $\boldsymbol{Q}_{FE}$ over the potential barriers unviable for practical THz field strengths. Due to such structural characteristics, a THz-induced transient polar phase can be observed below the equilibrium critical temperature of ~36 K in quantum paraelectric SrTiO$_3$[30,43], while the polarization in ferroelectric BaTiO$_3$ alters only by ~10% at room temperature under similar THz excitations[33].

**Simulations of the collective dynamics of polar skyrmions**

Here, we explore single-cycle THz pulse-induced ultrafast dynamics of the polar skyrmions stabilized in (PbTiO$_3$)$_{16}$/(SrTiO$_3$)$_{16}$ superlattices. **Fig. 1a** illustrates the schematic of the THz excitation phenomena of the system, adapted from the results of dynamical phase-field simulations within the Landau-Ginzburg-Devonshire (LGD) theory framework (see Methods). The system is characterized by the in-plane periodic distribution of individual skyrmions within the PbTiO$_3$ layers, in between which continuously rotating polarization vectors form the skyrmion walls. Built on the Last phonon mode of PbTiO$_3$, multiple collective dynamical modes are hosted in the skyrmions, and their corresponding real-space motions primarily consist of coherent wobbling of the skyrmion walls, accompanied by oscillations of the cores along the out-of-plane direction (see Supplementary Movie S1). These features are qualitatively similar to the collective modes previously observed in polar vortices.[24] As shown in **Fig. 1b**, the simulated frequency-domain response spectra indicate a number of well-defined collective modes below ~2 THz, overlapping with the experimental THz pulse spectrum (generated via the LiNbO$_3$ optical rectification method). These low-lying collective modes (with mode coordinate $Q_{Sk}$) provide multiple coupling channels with external THz electric fields. Compared with architype ferroelectrics, the presence of periodic skyrmion walls weakens the long-range dipole correlation, cancels out the macroscopic polarization[44], and leads to a flattened energy landscape rather than a double-well potential with respect to $Q_{Sk} = 0$[7] (see Supplementary Fig. S2). These factors facilitate the driven motions of $Q_{Sk}$ into large magnitudes; once $Q_{Sk}$ gains nonvanishing values macroscopic symmetry-breaking occurs in the skyrmions, leading to the emergence of a transient polar phase (**Fig. 1a**). Besides, the collective dynamics of the skyrmions marginally shift in frequencies with temperature until a topological phase transition sets in above 500 K (**Fig. 1b**), thus providing a remarkable THz coupling efficiency across a broad temperature range. These unique dynamical properties are rooted in the topological protection attribute of the skyrmion superlattice, distinct from its components PbTiO$_3$ and SrTiO$_3$.

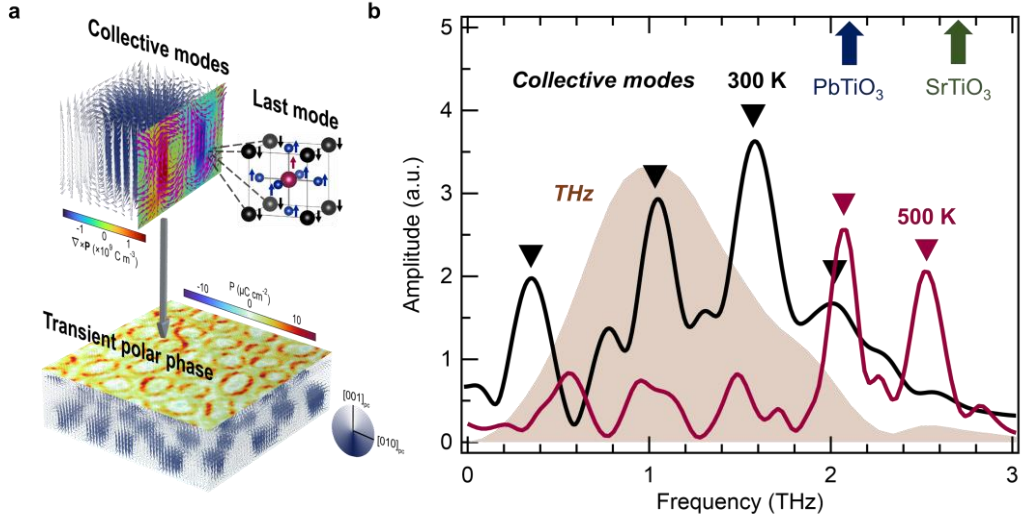

**Fig. 1. Simulated collective dynamics of the polar skyrmions.**
**a** Dynamical phase-field simulation of the collective dynamics of the polar skyrmions in $(PbTiO_3)_{16}/(SrTiO_3)_{16}$ superlattices. *Upper left*: Static configuration of a single skyrmion and the vertical cross-sectional view of the vorticity of the polarization vectors. The arrows denote the motion vectors of ions (pink) or local polarization vectors (white-blue). *Upper right*: Schematic of the Last phonon mode in a single unit cell which constitutes the building block of the collective modes. *Lower*: Excited state of the skyrmions forming a transient polar phase, together with the horizontal cross-sectional view of the polarization magnitude. **b** Fourier-transformed spectra of the time-domain polarization response of the polar skyrmions upon THz electric field excitation, simulated for 300 K (black) and 500 K (red). The spectrum of the experimental THz pulse is overlaid in the background. The soft-mode frequencies of $SrTiO_3$ and $PbTiO_3$ at 300 K are indicated with arrows.

**THz-induced phase transition in polar skyrmions**

To measure the predicted dynamical mechanisms, we performed ultrafast THz-pump, second harmonic generation (SHG) probe and optical Kerr effect probe spectroscopies. As schematically shown in **Fig. 2a** (see more details in Methods and Supplementary Fig. S3), the THz pump and 800 nm probe were collinearly directed onto the sample at normal incidence, with the THz electric field oriented along the [100] direction (in the pseudocubic setting). The 400 nm SHG signals were collected in a near-backscattering direction, sensitive to the THz field-induced SHG (TFISH) response arising from the in-plane symmetry breaking[45]. The polarization angles of the 800 nm and 400 nm lights (denoted as $\alpha$ and $\varphi$, respectively) were varied relative to the THz field direction; for convenience, we define the parallel and perpendicular directions as *p*- and *s*-, respectively. The measurements included TFISH polarimetry scans (varying $\alpha$ and/or $\varphi$ angles) at a fixed pump-probe delay time and delay time scans under the *p*-in/*p*-out ($\alpha = \varphi = 0°$) and *s*-in/*p*-out ($\alpha = 90°/\varphi = 0°$) probing conditions. In addition to skyrmion-hosting $(PbTiO_3)_{16}/(SrTiO_3)_{16}$ superlattices, single *c*-domain $PbTiO_3$ thin films and $SrTiO_3$ single crystals were also investigated as model systems of architype (incipient) ferroelectrics.

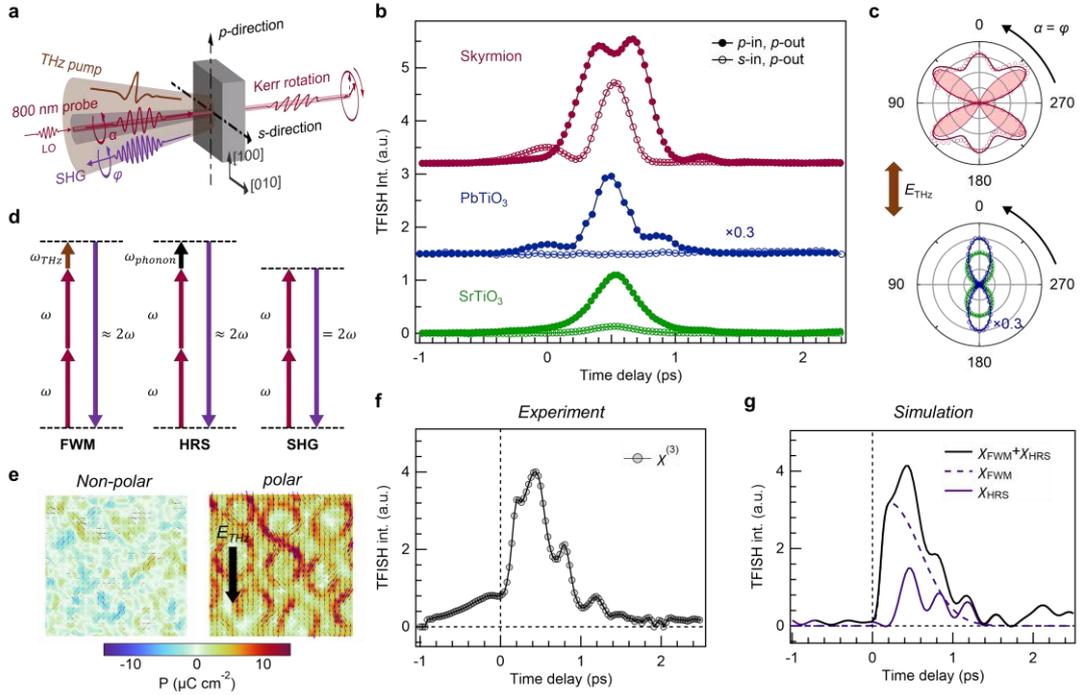

**Fig. 2. THz field-induced second harmonic generation (TFISH) in the polar skyrmions.**
**a** Schematic of the ultrafast THz measurement geometry and the definition of the principal axes. LO: local oscillator. **b** Time-domain TFISH response spectra for the polar skyrmions, $PbTiO_3$ thin films and $SrTiO_3$ single crystals. The *p*-in/*p*-out spectrum of $PbTiO_3$ is multiplied by 0.3 for clarity. **c** TFISH polarimetry plots measured at a delay time of ~0.6 ps for the three systems. The $\chi^{(2)}$ contribution extracted from the model fitting is highlighted by shadow. **d** Schematic of the three distinct TFISH mechanisms: four-wave mixing (FWM), hyper-Raman scattering (HRS) and second harmonic generation (SHG). **e** Spatial distribution of the in-plane polarization of the polar skyrmions before and after the THz field excitation, obtained from dynamical phase-field simulations. **f** Model-fitted $\chi^{(3)}$ contribution to the *p*-in/*p*-out TFISH spectra of the polar skyrmions. **g** Decomposition of the FWM and HRS responses to the $\chi^{(3)}$ contribution based on the dynamical phase-field simulations. All the experiment results here were measured at ~300 K.

**Fig. 2b** presents the time-domain TFISH response spectra of the three systems measured at 300 K. These spectra generally follow the envelope of the squared THz waveform, reaching peak responses at a delay time of approximately 0.6 ps; note that the dip in the *p*-in/*p*-out spectra of the polar skyrmions at 0.6 ps is due to their oscillatory behavior (to be discussed later). For PbTiO$_3$ and SrTiO$_3$, large TFISH responses are observed under the *p*-in/*p*-out conditions while weak to no responses are detected under the *s*-in/*p*-out conditions, indicating a pronounced anisotropy. By contrast, the polar polar skyrmions exhibit comparable TFISH responses in both cases. Further information on the TFISH anisotropy is obtained from the polarimetry patterns measured at the peak delay time. **Fig. 2c** presents the results of the $\alpha/\varphi$-coupled scan, equivalent to rotating samples under the $\alpha = \varphi = 0°$ conditions (see $\alpha$-scan results in Supplementary Fig. S4). The patterns for PbTiO$_3$ and SrTiO$_3$ show similar twofold symmetries with two lobes parallel to the THz electric field. Apart from these two lobes, the patterns for the polar skyrmions manifest an additional set of four lobes and altogether resemble a six-fold-like symmetry. Such six-fold-like patterns are unusual given the pseudocubic crystal symmetry of PbTiO$_3$/SrTiO$_3$ superlattices and suggest complex TFISH mechanisms of the skyrmion system.

Phenomenologically, the TFISH response of materials is the total contribution from the third-order ($\chi^{(3)}$) and second-order ($\chi^{(2)}$) optical nonlinearities, mediated by three mechanisms: four-wave mixing (FWM), Hyper-Raman scattering (HRS) and SHG (**Fig. 2d**) [46–48]. FWM is a three-photon process involving two probe photons and one THz pump photon, generating a photon at $2\omega_{probe} \pm \omega_{THz}$. HRS refers to the interaction between the probe light and a material's phonon, generating a photon at $2\omega_{probe} \pm \omega_{phonon}$. SHG is a two-photon process of the probe light, generating a photon at exactly $2\omega_{probe}$. Both FWM and HRS manifest as sidebands near SHG, but practically are spectrally indistinguishable in TFISH measurements due to the large bandwidth of femtosecond probe pulses (~30 nm). The FWM response arises from the intrinsic electronic anharmonicities and is not restricted to non-centrosymmetric materials, as confirmed here for all three systems. Their observed TFISH peaks in the *p*-in/*p*-out spectra and twofold polarimetry patterns (**Fig. 2b,c**) indicate a dominant $\chi^{(3)}$ response along the THz field direction. By contrast, the HRS response only occurs when the material's phonon frequency overlaps with the probe light. Due to the ineffective coupling of their soft phonons with the THz field at 300 K (**Fig. 1b**), SrTiO$_3$ and PbTiO$_3$ exhibit marginal HRS responses, as indicated by the non-oscillatory *p*-in/*p*-out TFISH spectra. For both systems, the weak TFISH responses measured under the *s*-in/*p*-out conditions (i.e., orthogonal to the THz field direction) also imply the absence of SHG responses, consistent with the fact that paraelectric SrTiO$_3$ has zero $\chi^{(2)}$ and the non-zero static $\chi^{(2)}$ of ferroelectric PbTiO$_3$ is not detected using the dynamic lock-in method.

The complex TFISH effects observed in the polar skyrmions suggest prominent HRS and SHG contributions. Due to the effective coupling between their collective modes with the THz field, the HRS responses are activated in the polar skyrmions to drive ***Q***$_{Sk}$ into large amplitudes, consequently breaking the centrosymmetry at equilibrium and inducing a transient polar phase with strong SHG response. The phase-field simulation results (**Fig. 2e**) indicate that local dipoles become ordered around the skyrmion walls, amounting to a net in-plane polarization along the THz field. This THz-induced transition is determined by the ***Q***$_{Sk}$ dynamics, leading to systematic temporal evolutions in the TFISH polarimetry patterns (see Supplementary Fig. S5). Furthermore, we quantify the FWM, HRS and SHG contributions from model fittings to the measured TFISH results (see Supplementary Note 1). As shown in **Fig. 2f**, the fitted $\chi^{(3)}$ contribution exhibits oscillations overlaid on a fast-decaying slope within ~2 ps. The two $\chi^{(3)}$ components, FWM and HRS, can be delineated based on the dynamical phase-field simulations (**Fig. 2g**). The HRS response is derived from the simulated time-dependent mean polarization values, while the electronic FWM response is approximated by a Gaussian profile, altogether closely matching the experiment results. The fitted $\chi^{(2)}$ SHG contribution shows no oscillatory behavior, in line with the transient polar order induced in the system (see Supplementary Fig. S6).

The aforementioned TFISH mechanisms are cross-validated by examining the THz field dependence (see Supplementary Fig. S7). A power factor of ~2 and ~1.75 is found for the *p*-in/*p*-out and *s*-in/*p*-out responses, respectively. The former power factor agrees with the $\chi^{(3)}$ contribution, whereas the latter suggests that the $\chi^{(2)}$ of the THz-induced phase is not necessarily proportional to THz field.

**Time-domain ultrafast dynamics of polar skyrmions**

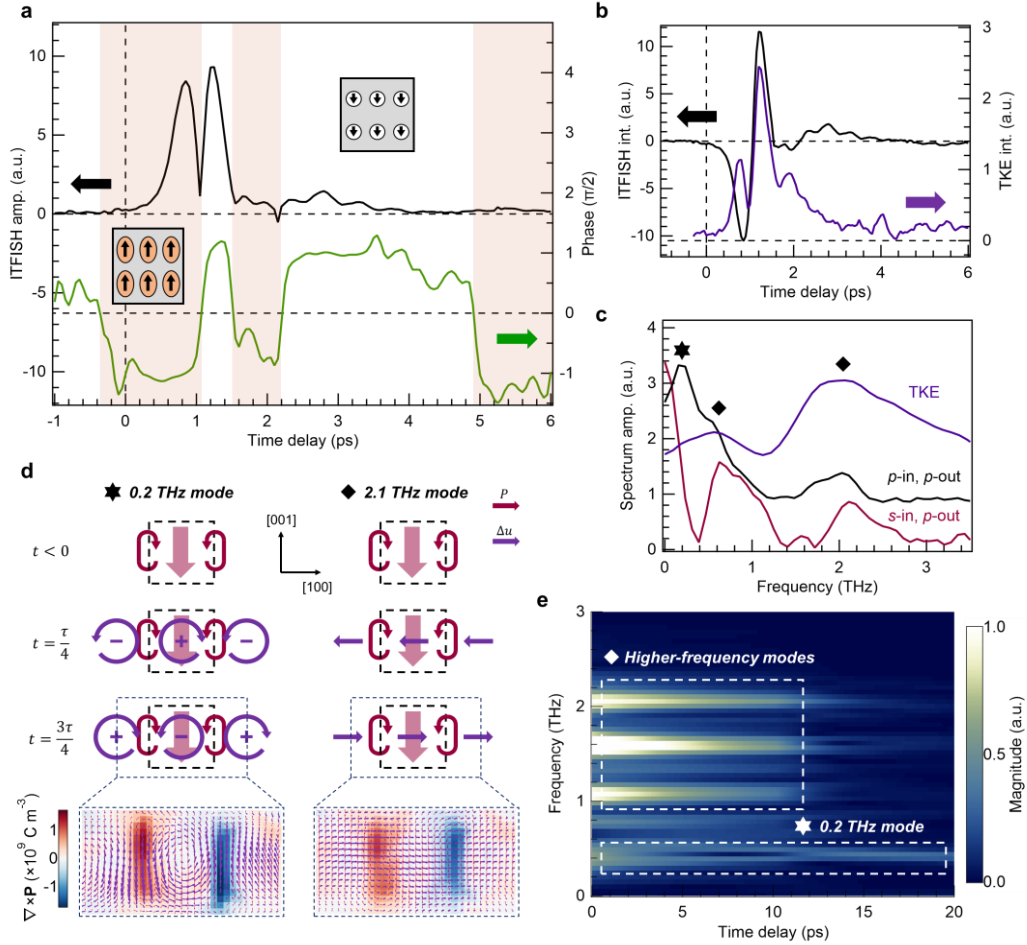

**Fig. 3. THz-induced time-domain dynamics of the polar skyrmions.**
**a** Time-domain spectra of the iTFISH amplitude (black) and phase (green) for the polar skyrmions, extracted from the interferometric measurements under the *p*-in/*p*-out condition. *Insets*: illustration of the transient dipole ordering. **b** Time-domain spectra of the phase-signed *p*-in/*p*-out iTFISH (black) and TKE (violet) responses for the polar skyrmions. **c** Fourier-transformed frequency spectra of the *p*-in/*p*-out iTFISH, *s*-in/*p*-out iTFISH, and the TKE measurement results. **d** Schematics of the spatial distributions of local polarization and motion vectors for the representative collective modes of the polar skyrmions. A skyrmion (dashed lines) consists of a quasi-single domain core (light red arrows) surrounded by walls (red arrows) where the polarization vectors rotate around the core. The corresponding phase-field simulated patterns are shown below. **e** Short-time Fourier transformation results of the phase-field simulated THz-induced response. All the experiment results here were measured at ~300 K.

We next focus on the ultrafast dynamics of the polar skyrmions to elucidate the mechanisms of the THz-induced polar phase. We incorporated interferometric TFISH (iTFISH) measurements by using a *β*-BaB$_2$O$_4$ single crystal as a local oscillator, allowing extraction of both the amplitude and phase information with enhanced sensitivities (see Methods and Supplementary Note 2) [49]. As shown in **Fig. 3a**, the iTFISH amplitude spectrum features two prominent peaks upon the main THz pulse, followed by oscillations over an extended period corresponding to the HRS responses of the collective modes. The iTFISH phase spectrum exhibits several π switching events, thus reflecting changes in the macroscopic polarization direction (**Fig. 3a** inset) that flip the signs of the nonlinear susceptibility tensors. These changes are not uniform during the initial ~6 ps period, suggesting a competition between multiple excited collective modes (**Fig. 1b**). The inherent connection with the collective dynamics underscores viable THz control over the macroscopic polar state of the system.

The phase-signed iTFISH signals ($I_{TFISH}\cos(\varphi)$) provide direct tracing of the $Q_{SK}$ mode coordinates, as shown in **Fig. 3b**. For comparison, we also measured the THz-induced Kerr effect (TKE), which are sensitive to changes in the linear optical susceptibility of the polar skyrmions[50]. Similar to TFISH, the TKE spectrum also shows a large initial electronic response and then a trail of oscillations after the main THz pulse. The frequency spectra of these oscillations are obtained from the Fourier transformation of the measured iTFISH and TKE spectra (**Fig. 3c**). They exhibit three main mode frequencies approximately at 0.2 THz, 0.6 THz and 2.1 THz, in remarkable agreement with the phase-field simulation results (**Fig. 1b**). Note that the 0.2 THz mode is observed only in the *p*-in/*p*-out TFISH signals, while the other modes appear in all three measurements. This difference in mode sensitivity is attributed to the symmetries of the collective modes, determined by their spatial polarization configurations.

The dynamical phase-field simulations further provide real-space microscopic insights into the skyrmion collective modes. Using a frequency decomposition method (see Supplementary Note 3), we extracted the spatio-temporal dynamics of local polarization (***P***) and motion vectors (***u***) for representative modes. **Fig. 3d** presents schematics of the 0.2 THz and 2.1 THz modes along with the corresponding phase-field simulated patterns (also see Supplementary Movie S2). The collective modes alter the polarization vectors' magnitude ($\Delta \boldsymbol{P}$), accompanied with atomic motions $\Delta \boldsymbol{u}$ (violet arrows). The 0.2 THz mode exhibits a circular pattern of atomic motions with alternating directions between adjacent skyrmions, thereby producing a temporal oscillation of vorticity ($\nabla \times \boldsymbol{u}$). Notably, this lowest-frequency mode appears to be in-plane symmetric, as the atomic motions cancel out between each pair of the adjacent circular patterns. By contrast, the 2.1 THz mode involves in-plane atomic motions with strong directionality along the THz field direction (the [100] crystal axis). This mode breaks the in-plane centrosymmetry, significantly contributing to both HRS and SHG responses. Therefore, the 2.1 THz mode is detected under both *p*-in/*p*-out and *s*-in/*p*-out conditions, whereas the 0.2 THz mode only produces an HRS signal detectable in the *p*-in/*p*-out condition. The skyrmion collective modes also exhibit different relaxation behaviors, as indicated by the short-time Fourier transform of the simulated THz response (**Fig. 3e**). The higher frequency modes decay rapidly within the first 10 ps, in contrast to the 0.2 THz mode which persists throughout the simulated period. This is consistent with the iTFISH results showing a fast relaxation of the oscillatory modes with high frequencies (**Fig. 3a**). The long-living 0.2 THz mode can be observed up to 80 ps (see Supplementary Fig. S8). Altogether, we conclude that while multiple collective modes of the polar skyrmions are excited to form a superimposed state, the fast-decaying higher-frequency modes dominate the emergence of the transient polar phase.

**Temperature-dependent behaviors**

The TFISH dynamics of the polar skyrmions exhibits unique temperature evolution behaviors. **Fig. 4a** presents their peak TFISH responses over a wide temperature range of ~4-500 K. The results for PbTiO$_3$ thin films are also included for comparison. Most notably, the TFISH responses of the polar skyrmions, measured under both the *p*-in/*p*-out (as has been attributed to the FWM/HRS contributions) and *s*-in/*p*-out (the SHG contribution) conditions, undergo an abrupt change at ~470 K reminiscent of a first-order phase transition. With increasing temperature until 470 K, the FWM/HRS contribution increases steadily, while the SHG contribution shows a moderate decreasing trend with several steps. Except for the overall intensities, the TFISH spectral features appear to be largely unchanged for both probing conditions (see Supplementary Fig. S9), corroborating the phase-field results (**Fig. 1b**). These observations of the polar skyrmions are qualitatively different from those of $PbTiO_3$, which essentially exhibits a consistent TFISH response under the *p*-in/*p*-out condition within the entire temperature range. The latter trend of $PbTiO_3$ can be expected given that its ferroelectric phase transition occurs at a much higher temperature (~760 K); for the same reason, its soft mode frequency remains to lie well beyond the THz excitation spectrum, thus resulting in negligible TFISH response under the *s*-in/*p*-out condition. Additionally, a surge in the (*p*-in/*p*-out) TFISH response signifying ferroelectricity is confirmed for $SrTiO_3$ at 5 K (**Fig. 2c**), again contrasting with the polar skyrmions[30].

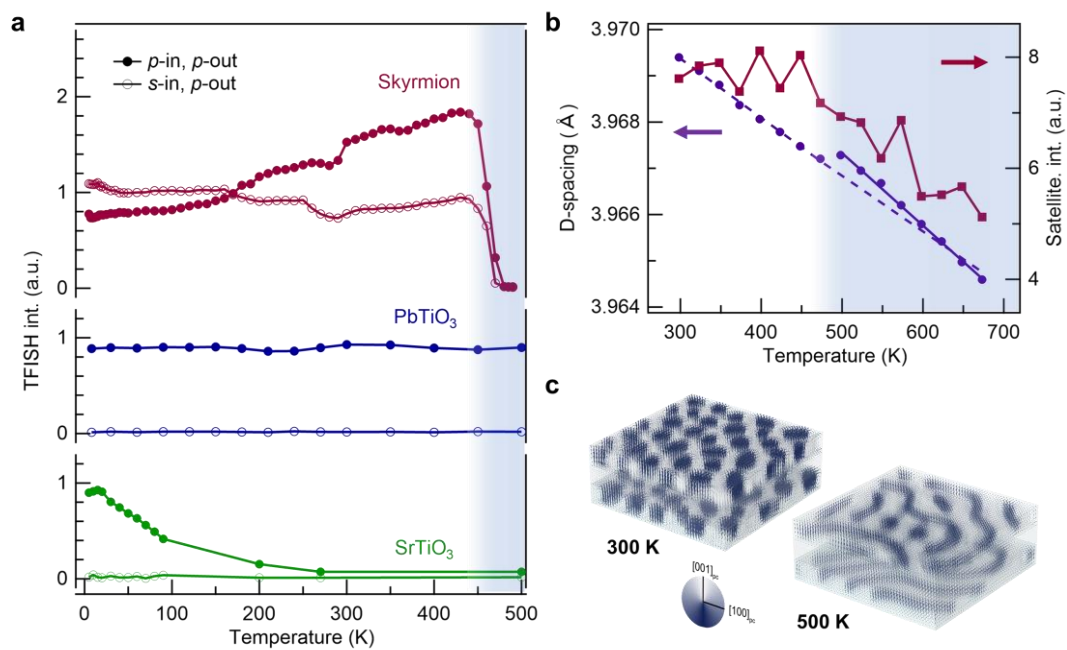

**Fig. 4. Temperature dependence of the THz-induced skyrmion dynamics.**
**a** Temperature evolution of the peak TFISH responses of the polar skyrmions, $PbTiO_3$ thin films and $SrTiO_3$ single crystals measured under the *p*-in/*p*-out and *s*-in/*p*-out conditions. **b** Temperature-dependent X-ray diffraction results of $(PbTiO_3)_{16}/(SrTiO_3)_{16}$ superlattices. The out-of-plane lattice parameters are extracted from the $\theta$-$2\theta$ scan patterns and the satellite peak intensities associated with the in-plane periodicity are extracted from the $\omega$-rocking curves. **c** An illustration of the near-470 K structural transition obtained from phase-field simulations.

The structural transitions of $(PbTiO_3)_{16}/(SrTiO_3)_{16}$ superlattices are revealed by temperature-dependent X-ray diffraction measurements within 300-670 K. As depicted in **Fig. 4b**, the out-of-plane lattice parameter exhibits a discernible kink around 470 K, and the linear thermal expansion coefficients slightly differ across the transition. Note, however, that such a transition is not accompanied with a complete loss of the in-plane ordered state, as evidenced by the preserved satellite diffraction intensities (see Supplementary Fig. S10). These observations corroborate the TFISH results and suggest a subtle structural transition path of the system, unidentified in previous reports of polar skyrmions. To reveal the likely polarization configurations above 470 K, we simulate the thermal effects by adjusting the substrate strain and Landau energy coefficients in the phase-field model. **Fig. 4c** presents the obtained polarization configuration for 500 K versus 300 K, illustrating much elongated, meandering dipole assemblies in deviation from the skyrmions. While further microscopic validations are warranted (for example, using ultrafast X-ray scattering techniques), the simulations indicate that the observed structural transition at 470 K can substantially shift the frequencies of the collective modes (**Fig. 1b**). The TFISH responses, through either HRS or related SHG mechanisms, are thereby quenched due to a disrupted coupling between the superlattice with the THz field.

**Discussion**

Here, we give a simplified account for the ultrafast THz excitation behaviors of the polar skyrmions. Established as an energetically frustrated yet spatially ordered system, the skyrmions essentially comprise a continuum of frozen states along the structural evolution path of $PbTiO_3$. The local polarization is fully developed at the skyrmion centers and highly suppressed near the walls, corresponding to local ferroelectric and paraelectric states, respectively. The paraelectric wall regions have been found to stabilize steady-state negative capacitance[7,8]. In terms of lattice dynamics, the continuum of states should encompass the temperature-dependent soft mode dynamics of $PbTiO_3$, which renormalizes into a set of resonant collective modes. The wall regions sustain considerable softening within the stability field of the skyrmions and thus are most susceptible to THz excitation. This characteristic originates from the topologically protected phase stabilities under THz field and renders the broad temperature range for an effective excitation, unsupported by architype (incipient) ferroelectrics including $PbTiO_3$ and $SrTiO_3$. The skyrmions are also dissimilar from chemically heterogenous materials with polar nanoregions, such as $KTaO_3$ and $Pb(Mg_{1/3}Nb_{2/3})O$ relaxors[51–54]. In the latter systems, dipolar relaxation dynamics is introduced by the randomly distributed polar nanoregions and moderately couples to THz fields at low temperatures. Furthermore, in the present case, the THz-induced transition is dominated by a strong collective mode at ~2.1 THz. This could potentially be tuned through modifying the superlattice structure or applying an external electric or stress field, by virtue of the structural robustness of the skyrmions.

In summary, we have observed the coherent excitations of the polar skyrmions driven by single-cycle THz pulses. Based on the TFISH/TKE measurements and dynamical phase-field simulations, we identify the connection between the spectrally complex macroscopic nonlinear optical responses with THz-excited collective modes of the skyrmions, highlighting the critical role of high-frequency collective modes in stabilizing the transient polar phase. The ultrafast dynamical behaviors of the skyrmions are stable against temperature, until they undergo a structural phase transition at ~470 K. All these observations posit the skyrmions, and likely also other topological polar systems, a testbed with both rich ultrafast dynamical phenomena and considerable application prospects in high-speed nonlinear electro-optic devices.

## Methods

### Sample fabrication

$[(PbTiO_3)_{16}/(SrTiO_3)_{16}]_8$ superlattices were fabricated with reflection high-energy electron diffraction (RHEED) assisted pulsed laser deposition (PLD, Anhui Epitaxy Technology). (001)-cut $(LaAlO_3)_{0.3}$-$(SrAl_{0.5}Ta_{0.5}O_3)_{0.7}$ (LSAT) single crystals were used as substrates. $PbTiO_3$, $SrTiO_3$ and $SrRuO_3$ ceramic targets were ablated by a 248 nm KrF excimer laser (COMPex205, Coherent) with a fluence of 1.0 J cm$^{-2}$. The bottom electrode layer of $SrRuO_3$ was grown at 680 °C under 10 Pa oxygen pressure. A 6 nm thick SrTiO$_3$ buffer layer, followed by a periodic sequence of PbTiO$_3$ and SrTiO$_3$ layers, was grown at 650 °C under 12 Pa oxygen pressure. Precise control of the unit-cell layer number was confirmed by RHEED intensity oscillations. For the two control samples, single-layer PbTiO$_3$ thin films were grown at 700 °C under 12 Pa oxygen pressure, and a commercial (001)-cut SrTiO$_3$ single crystal was used.

For the skyrmion samples in EFISH measurements, the same superlattice structure of [(PbTiO$_3$)$_{16}$/(SrTiO$_3$)$_{16}$]$_8$ was grown directly on LSAT substrates without a SrRuO$_3$ bottom electrode layer, and the deposition parameters were finely adjusted to optimize the growth quality. After the growth, several pairs of ~150 nm sputtered Pt coplanar electrodes with ~9 μm gap were obtained using photolithography lift-off to facilitate the application of an in-plane electric field.

**Structural characterizations**

Thin-film X-ray diffraction was carried out with a Malvern PANalytical Empyrean diffractometer employing Cu K$_{\alpha 1}$ radiation (wavelength: 1.5406 Å). The measurements included $\theta$–2$\theta$ specular scans and $\omega$-rocking scans. The superlattice samples were placed in a vacuum chamber and heated from room temperature to 700 K. The lattice constants were calculated from the zero-order superlattice diffraction peak in the $\theta$–2$\theta$ scan patterns. The satellite diffraction intensities were derived from the Gaussian fitting results of the satellite peaks near $\omega = 21.5°$ and $\omega = 24.1°$ in the rocking curves.

**Ultrafast THz spectroscopies**

A Ti:sapphire femtosecond laser amplifier (Spitfire Ace, Spectra-Physics) generated 35 fs pulses with a central wavelength of 800 nm, a repetition rate of 1 kHz, and a single-pulse energy of 7 mJ. The 95% laser output was employed to generate THz radiation from a LiNbO$_3$ crystal by the tilted wavefront technique[55]. The THz radiation was focused onto the sample at normal incidence with the electric field along the [100] direction. The remaining weak laser output was used as a probe beam to generate either SHG or optical Kerr signal from the sample. The probe pulses were focused on the sample surface at normal incidence, the 400 nm SHG light was collected in a near-backscattering direction, and the Kerr signal was detected by a balanced detector. For the TFISH measurements, the polarization direction of the probe beam was controlled by a half-wave plate and formed an angle $α$ relative to the THz field, while a polarizer was used to select the SHG light with a polarization direction at an angle $φ$ to the THz field. The SHG light was then delivered through several bandpass filters and detected using a photomultiplier tube. An optical chopper modulated the pump THz beam at 500 Hz. The measured signal was demodulated with the same frequency by a lock-in amplifier, ruling out static SHG signals that would generally exist in ferroelectric systems such as $PbTiO_3$. A continuous-flow He cryostat with sapphire windows was used to vary the sample temperature from 4 K to 500 K. For the interferometric TFISH measurements, a $β$-$BaB_2O_4$ (BBO) single crystal was put into the probe line to generate the LO pulses colinearly in the orthogonal polarization direction. A calcite single crystal was used to compensate the optical path length difference between the probe and the LO light beams. A pair of fused-silica wedges (2.8°) served as the phase modulator and were adjusted with a piezoelectric displacement stage. Finally, the polarization direction of the probe and the LO beams were rotated to be parallel using a dual wavelength waveplate. For the TKE measurements, the polarization direction of the probe beam was fixed at $α = 45°$ relative to the THz field.

**Dynamical phase-field simulations**

A phase-field model for the superlattice system was constructed to simulate the THz-induced ultrafast dynamics of the skyrmions based on a finite-element method solver (COMSOL Multiphysics). The model consisted of 5 periods of alternating $PbTiO_3$ and $SrTiO_3$ layers, each with a thickness of 6.4 nm (approximately 16 unit cells), atop a 10 nm substrate layer. Along the [100] and [010] directions, the model size was $50×50$ $nm^2$ (approximately $125×125$ unit cells). For the top and bottom (001) surfaces, short-circuit electrical boundary conditions were applied. For the two pairs of in-plane (100)/(010) surfaces, periodic boundary conditions were applied. At the bottom surface of the substrate, the mechanical displacements were fixed, corresponding to -1.2% biaxial strains induced by the lattice mismatch between $PbTiO_3$ and LSAT, while the top surface was mechanically free. The superlattice layers were assumed to be fully constrained to the substrate.

The free-energy functional $F$ is defined in the form:

$$F = \int (E_{land} + E_{elec} + E_{elas} + E_{grad})dV \qquad (2)$$

where $E_{land}, E_{elec}, E_{elas}, E_{grad}$ correspond to the Landau, electrostatic, mechanical and polarization gradient energy densities, respectively. In the model, these energetic terms were evaluated using three sets of field variables, including polarization (***P***), mechanical displacement (***u***), and electric potential ($\varphi$). The spatiotemporal dynamics of the model are governed by a set of coupled equations of motion:

$$\mu \ddot{P}_i + \gamma \dot{P}_i = -\delta F/\delta P_i \qquad (3)$$

$$\rho \ddot{u}_i + \beta \rho \dot{u}_i = -\delta F/\delta u_i \qquad (4)$$

$$0 = -\delta F/\delta \varphi \qquad (5)$$

where $\mu$, $\gamma$, $\rho$ and $\beta$ are the effective polarization mass density, polarization damping coefficient, mass density and elastic damping coefficient, respectively. Eq. (3) is the second-order time-dependent LGD equation which determines the polarization dynamics, and Eq. (4) governs the elastodynamics. Eq. (5) is Poisson's electrostatic equation which neglects complex electrodynamics related to charge carrier transport, since the intrinsic carrier density is low in the skyrmion system and the THz excitation does not have large enough energy to populate free electron/hole carriers. For the superlattice layers, all three equations were employed, while for the substrate layer only Eq. (4) was enabled.

The steady-state configuration of the skyrmions was first obtained by setting both $\mu$ and $\rho$ to zero (thus nullifying the mass inertial effects) and evolving from small random initial polarization values (<0.0007 C m$^{-2}$) in the superlattice layers. For simulating THz excitation of the skyrmions, the $\mu$ and $\rho$ were turned on and the steady-state polarization configuration was set as the initial values. An additional electric field component was included in the equations, which had the same waveform as the electro-optic sampling results; subject to this virtual THz pulse, the evolution of the model was calculated within a time range of 0-40 ps. To best fit the experimentally observed TFISH response, the $\mu$ was optimized to a value of 7.53×10$^{-18}$ kg m$^3$ A$^{-2}$ s$^{-2}$. More details about the phase-field model can be found in [26] and [56].

**Data availability**

The experimental and simulation data presented in this article are available from the corresponding author upon request.

**Code availability**

The data analysis codes used to produce all the results presented in this article are available from the corresponding authors upon request.

**Acknowledgments**

This work was financially supported by Basic Science Center Project of National Natural Science Foundation of China (NSFC) under grant No. 52388201, NSFC grants No. U24A2009, No. 52150092, No. 11974414, National Key Basic Research Program of China under Grant No. 2020YFA0309100, and by Beijing Municipal Natural Science Foundation under Grant No. JQ24011 and Z240008. A portion of this work was carried out at the Synergetic Extreme Condition User Facility (SECUF, https://cstr.cn/31123.02.SECUF).

**Author contributions**

W.L. and X.W. performed the THz spectroscopies measurements. P.P., X.W. and J.L. developed the experimental setup. S.W. fabricated the samples. W.L. and Q.L. performed the phase-field simulations. H.H. and J.M. performed the XRD measurements. J.-M.L., J.-F.L. and C.-W.N. discussed the results. W.L. and Q.L. wrote the manuscript with inputs from all the co-authors. Q.L. conceived and supervised the project.

**Competing interests**

The authors declare that they have no competing interests.

# Supplementary Materials

# Broad-temperature-range ultrafast terahertz excitation of collective dynamics in polar skyrmions

Wei Li[a], Sixu Wang[a], Pai Peng[b,c], Haojie Han[a], Xinbo Wang[b]*, Jing Ma[a], Jianlin Luo[b], Jun-Ming Liu[d], Jing-Feng Li[a], Ce-Wen Nan[a], Qian Li[a]*

[a] *State Key Laboratory of New Ceramics and Fine Processing, School of Materials Science and Engineering, Tsinghua University, Beijing 100084, China*

[b] *Beijing National Laboratory for Condensed Matter Physics, Institute of Physics, Chinese Academy of Sciences, Beijing 100190, China*

[c] *State Key Laboratory of Low Dimensional Quantum Physics, Department of Physics, Tsinghua University, Beijing 100084, China*

[d] *Laboratory of Solid State Microstructures and Innovation Center of Advanced Microstructures, Nanjing University, 210093, Nanjing, China*

* Corresponding author: (QL) qianli_mse@tsinghua.edu.cn; (XW) xinbowang@iphy.ac.cn

**Supplementary Notes**

**S1 Fitting procedure for the TFISH results**

The TFISH results are fitted considering the contribution from the four-wave mixing (FWM), Hyper-Raman scattering (HRS), and second harmonic generation (SHG). The total TFISH response is expressed as:

$$I_{TFISH}^{2\omega} = I_{FWM}^{2\omega} + I_{HRS}^{2\omega} + I_{SHG}^{2\omega}$$
$$\propto \left(\chi_{FWM}^{(3)} E_{THz} E_{probe}^2\right)^2 + \left(\chi_{HRS}^{(3)} E_{phonon} E_{probe}^2\right)^2 + \left(\chi_{SHG}^{(2)} E_{probe}^2\right)^2 \quad (S1)$$

where the $\chi_{FWM}^{(3)}$, $\chi_{HRS}^{(3)}$ and $\chi_{SHG}^{(2)}$ are the nonlinear susceptibilities for the FWM, HRS and SHG processes, respectively. $E_{THz}$ and $E_{probe}$ are the field strength of the THz and 800 nm probe light beams. $E_{phonon}$ describes the effective field associated with a phonon driven by the THz pump. Note that for the THz field strengths used here (<1 MV cm$^{-1}$), the phonon nonlinearity of the skyrmion collective modes can be ignored and thus $E_{phonon} \propto E_{THz}$ (see Supplementary Fig. S11).

For the polarimetry measurements, the nonlinear susceptibilities are considered in their full tensor forms. Skyrmions exhibit complex topological polarization configuration, making the space-group symmetry quite unclear. Here, we assume that the SHG response $\chi_{SHG}^{(2)}$ obeys a *4mm* point group symmetry known for tetragonal-phase PbTiO$_3$. We define the coordinate system as the *x*, *y*, *z* axes corresponding to the [100], [010] and [001] crystal axes in order (the *x* axis corresponds to the THz direction), leading to a $\chi_{SHG}^{(2)}$ tensor form:

$$\chi_{SHG}^{(2)} = \begin{pmatrix} \chi_{11}^{(2)} & \chi_{13}^{(2)} & \chi_{13}^{(2)} & 0 & 0 & 0 \\ 0 & 0 & 0 & 0 & 0 & \chi_{35}^{(2)} \\ 0 & 0 & 0 & 0 & \chi_{35}^{(2)} & 0 \end{pmatrix} \quad (S2)$$

The FWM and HRS responses are not explicitly dependent on symmetry since they are third-order nonlinearities. According to the experiment results of the TFISH spectra, we found that these responses are dominant along the THz field direction (for all materials), indicating that most elements of the tensors to be essentially zero except for a uniaxial coefficient $\chi_{111}^{(3)}$. Moreover, the two responses are indistinguishable from a single fit to the TFISH polarimetry. We therefore fit the effective (FWM and HRS) third-order nonlinear susceptibility $\chi^{(3)}_{eff}$ as:

$$\chi^{(3)}_{eff} = \begin{pmatrix} \chi^{(3)}_{111} & 0 & 0 & 0 & 0 & 0 \\ 0 & 0 & 0 & 0 & 0 & 0 \\ 0 & 0 & 0 & 0 & 0 & 0 \end{pmatrix} \quad (S3)$$

Note that the tensor is expressed in a two-dimensional form since the THz field is strictly along the $x$ axis. For the $\alpha$-scan measurements (see Supplementary Fig. S12), $\varphi$ are kept at either 0° ($p$-out) or 90° ($s$-out), leading to an angle-dependent TFISH intensity formula as following:

$$I^p_{exp}(\alpha) \propto \left(\chi^{(2)}_{11} E^2_{THz} \sin^2\alpha + \chi^{(2)}_{13} E^2_{THz} \cos^2\alpha\right)^2 + \left(\chi^{(3)}_{111} E^3_{THz} \cos^2\alpha\right)^2 \quad (S4)$$

$$I^s_{exp}(\alpha) \propto \left(\chi^{(2)}_{35} E^2_{THz} \sin\alpha \cos\alpha\right)^2 \quad (S5)$$

For the $\alpha/\varphi$-coupled scan measurements, $\alpha$ and $\varphi$ are adjusted synchronously to keep $\alpha = \varphi$. The corresponding TFISH intensity can be expressed as:

$$I^c_{exp}(\alpha) = I^p_{exp}(\alpha) \cos\alpha + I^s_{exp}(\alpha) \sin\alpha \quad (S6)$$

The coefficients are assigned with trial values and then optimized using a standard Newton's method by minimizing the loss function. We used $L2$ loss function as the error function:

$$L = \sum_i \left(I_{fit}(\alpha_i) - I_{exp}(\alpha_i)\right)^2 \quad (S7)$$

where $\alpha_i$ is the polarization angle at the $i^{th}$ data point, $I_{fit}$ is the theoretical SHG intensity, and $I_{exp}$ is the measured SHG intensity. The fitted nonlinear susceptibilities at the maximum THz field are:

$$\chi^{(2)}_{SHG} = \begin{pmatrix} 55 & 55 & -1 & 0 & 0 & 0 \\ 0 & 0 & 0 & 0 & 0 & 54.9 \\ 0 & 0 & 0 & 0 & 54.9 & 0 \end{pmatrix} \text{arb. units} \quad (S8)$$

$$\chi^{(3)}_{eff} E_{THz} = \begin{pmatrix} 40.2 & 0 & 0 & 0 & 0 & 0 \\ 0 & 0 & 0 & 0 & 0 & 0 \\ 0 & 0 & 0 & 0 & 0 & 0 \end{pmatrix} \text{arb. units} \quad (S9)$$

For the time-domain measurements, we measured the time-domain polarimetry data composed of $\alpha$-scan and $\alpha/\varphi$-coupled scan results at each time delay $t$ (see Supplementary Fig. S7). Fitting to this data allows the extraction of time-dependent nonlinear susceptibilities $\chi^{(2)}_{SHG}(t)$ and $\chi^{(3)}_{eff}(t)$. The results are shown in main text Fig. 2f and g and Supplementary Fig. S8.

The $\chi^{(3)}_{eff}$ responses are further delineated into FWM and HRS components based on phase-field simulations. FWM originates from the electronic responses of the materials, and can be approximated by a Gaussian profile

$$\chi^{(3)}_{FWM}(t) = \frac{A}{\sigma\sqrt{2\pi}}\exp\left(-\frac{(t-t_0)^2}{2\sigma^2}\right)\exp\left(-\frac{t-t_1}{\tau}\right) \qquad (S10)$$

with the second term as a decay correction. The HRS response is proportional to the coordinate of the excited phonon, thus related to the expected value of polarization:

$$\chi^{(3)}_{HRS}(t) \propto Q_{phonon}(t) \propto \langle \boldsymbol{P}\rangle(t)$$

We calculated the mean polarization values $\langle \boldsymbol{P}\rangle(t) = \oiiint \boldsymbol{P}(x,y,z,t)dV$ across the entire PbTiO$_3$ region in the phase-field simulations. Finally, a simulation of $\chi^{(3)}_{eff}$ was completed by calculating $\chi^{(3)}_{eff}(t) = \chi^{(3)}_{FWM}(t) + \chi^{(3)}_{HRS}(t)$.

## S2 The calculation of the iTFISH amplitudes and phases

In our experiment setup, the 800 nm probe light and the local oscillator (LO) propagate colinearly before incident normally onto the sample. The outgoing 400 nm light is composed of the TFISH response, the reflected LO, and the interference of the two. Furthermore, due to the dynamical lock-in setup, the reflected LO is filtered out since the signal is not modulated by the chopper. In this case, the total response can be expressed as:

$$I = I_{TFISH} + \sqrt{I_{TFISH}I_{LO}}\cos(\Delta\varphi) \qquad (S11)$$

where $I_{TFISH}$ is the TFISH response, $I_{LO}$ is the LO intensity, and $\Delta\varphi$ is the phase difference between the two sources. We adjusted $\Delta\varphi$ using a pair of wedges made of fused silica, and the corresponding iTFISH signals were collected with PMT. As shown in Fig. S13, a well-defined trigonometric dependency was observed, corresponding to the second item in Eqn. S1. Note that in order to validate the experiment setup, here the TFISH signal is far smaller than the LO intensity, and $I \approx \sqrt{I_{TFISH}I_{LO}}\cos(\Delta\varphi)$.

Therefore, the iTFISH change oscillates between $\pm 1$. When measuring the time-domain spectra of polar skyrmions, the intensities of TFISH and LO are more balanced. In order to obtain the time-domain iTFISH signals of polar skyrmions, $\Delta\varphi$ was adjusted to a set of values from $-\pi$ to $\pi$, with each value doing a standard time-domain pump-probe scanning. The data was then rearranged into the iTFISH intensity of each time delay as a function of $\Delta\varphi$, that is, the point-wise experiment data of Eqn. S1. The data was then fitted to the time-domain expression of Eqn. S11:

$$I(t) = I_{TFISH}(t) + \sqrt{I_{TFISH}(t)I_{LO}} \cos(\Delta\varphi(t)) \tag{S12}$$

to determine the two independent variables $I_{TFISH}(t)$ and $\Delta\varphi(t)$, as is termed the amplitudes and the phases of the iTFISH signal in the main text. Note that $I_{TFISH}(t)$ was considered to be non-negative and the fitted phases $\Delta\varphi(t)$ were restricted to the $[-\pi, \pi]$ range.

## S3 Isolation of certain frequency components in phase-field simulation

To determine the spatial and temporal characteristics of specific collective phonon mode, an isolation method based on Fourier transformation was applied. A set of phase-field simulations were performed to obtain the spatially resolved time-domain evolution of the polarization vectors $\boldsymbol{P}(x,y,z,t)$. We first obtained the eigenfrequencies of the collective modes by calculating the mean polarization values $\langle\boldsymbol{P}\rangle(t) = \oiiint \boldsymbol{P}(x,y,z,t)dV$. Fourier transformations were performed on the spectra to obtain $\langle\boldsymbol{P}\rangle(\omega)$. By inspecting the frequency-domain spectra of the polarization along THz field $\langle\boldsymbol{P}_{[100]}\rangle(\omega)$, several eigenfrequencies $\omega_0$ were selected. We next performed Fourier-transformation point-by-point on $\boldsymbol{P}(x,y,z,t)$ to obtain $\widetilde{\boldsymbol{P}}(x,y,z,\omega)$ in its complex form. Inverse Fourier-transformations were then applied directly on $\widetilde{\boldsymbol{P}}(x,y,z,\omega_0)$ to calculate $\boldsymbol{P}_0(x,y,z,t)$ which correspondingly reflect the time-domain dynamics of the specific frequency component $\omega_0$. Note that this method cannot assure the decomposition of atomic motions into exactly the eigenmodes, but a numerical calculation result of the general characteristic of the dominant phonon mode at the frequency.

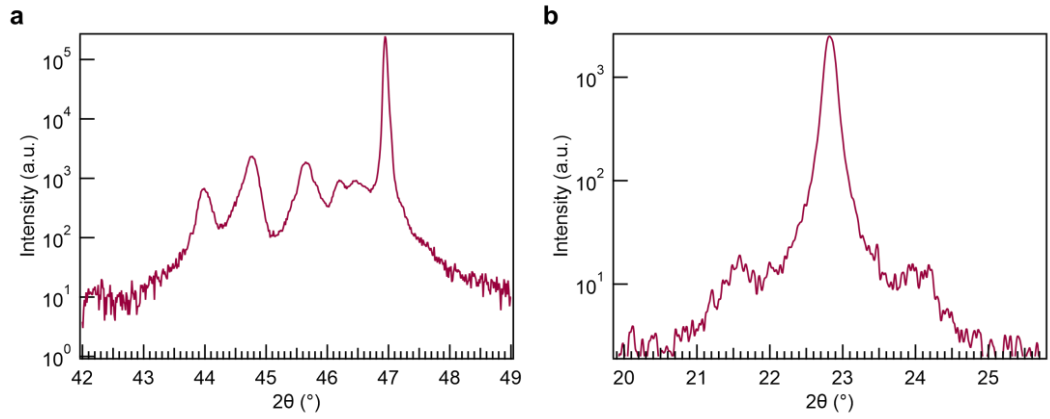

**Fig. S1. X-ray diffraction result of the skyrmion sample**
**a** X-ray intensity-$2\theta$ pattern of the polar skyrmions. **b** $\omega$-rocking scan result of the polar skyrmion near its 2nd superlattice peak in (**a**) around 44.8°.

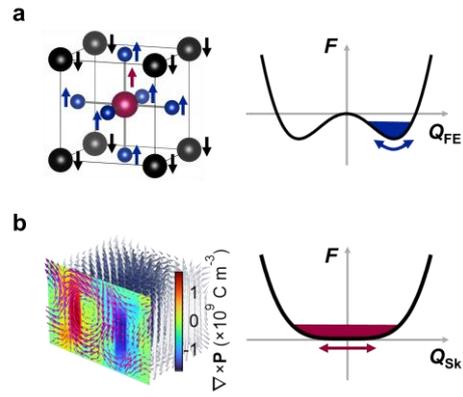

**Fig. S2. Comparison of the collective modes and the ferroelectric soft mode.**
Illustrations of **a** the PbTiO$_3$ soft mode and **b** the skyrmion collective modes, along with the free-energy profiles for both types of modes. The arrows denote the motion vectors of ions or local polarization vectors.

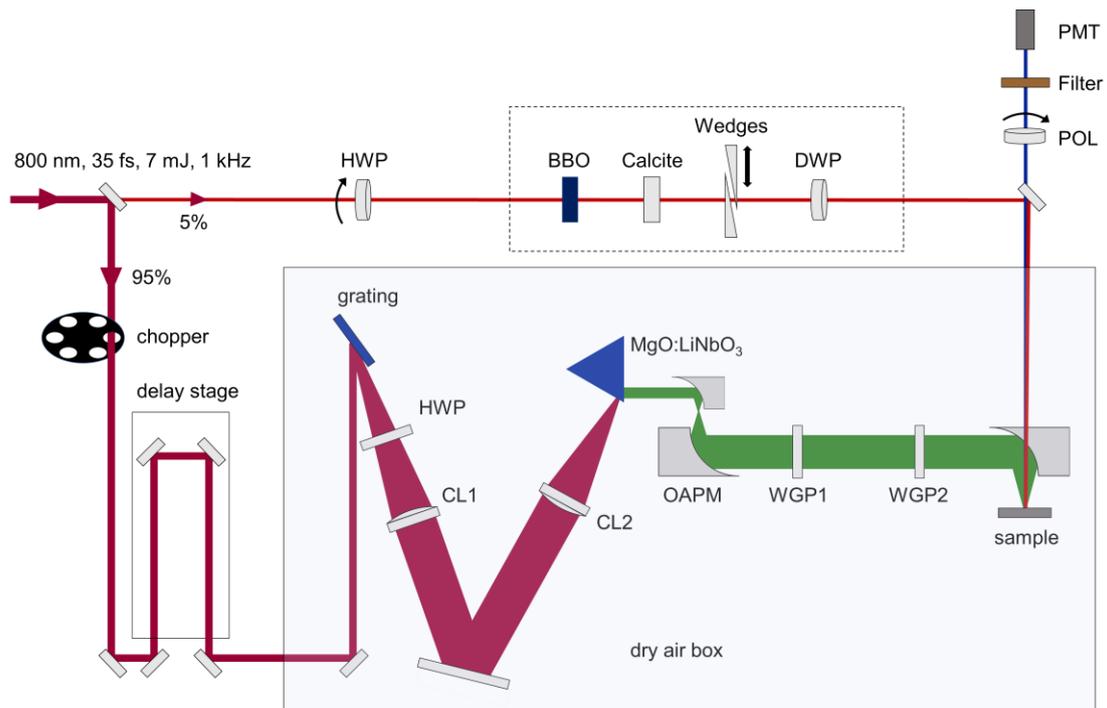

**Fig. S3. Illustration of the TFISH experiment setup**

An illustration of the TFISH experiment setup. HWP: half-wave plate; QWP: quarter-wave plate; CL: plano-convex cylindrical lens; OAPM: off-axis parabolic mirror; WGP: terahertz wire grid polarizer; PMT: photomultiplier tube.

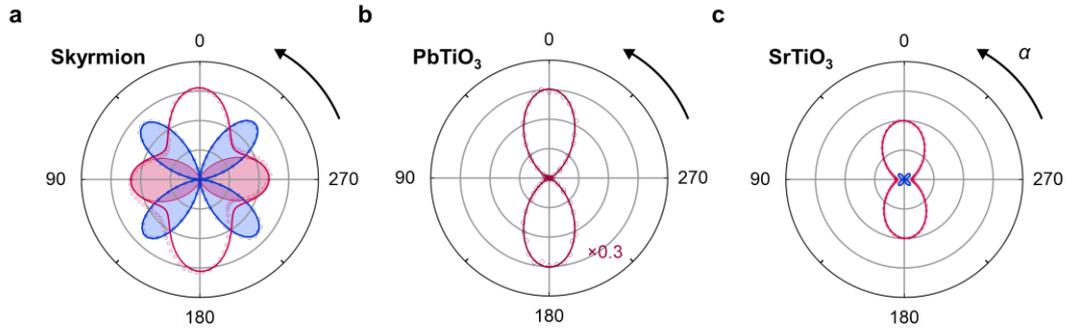

**Fig. S4. α-scan result of the polar skyrmions, PbTiO₃ and SrTiO₃**
SHG polarimetry at the TFISH peak around 1.7 ps time delay for (**a**) the polar skyrmions, (**b**) the PbTiO₃ thin film and (**c**) SrTiO₃ single crystal, respectively. The data was measured by rotating incident angle $\alpha$ while fixing $\varphi$ at 0° (red) and 90° (blue). For the polar skyrmions results, the $\chi^{(2)}$ contribution extracted from the fitting result is highlighted with shadows.

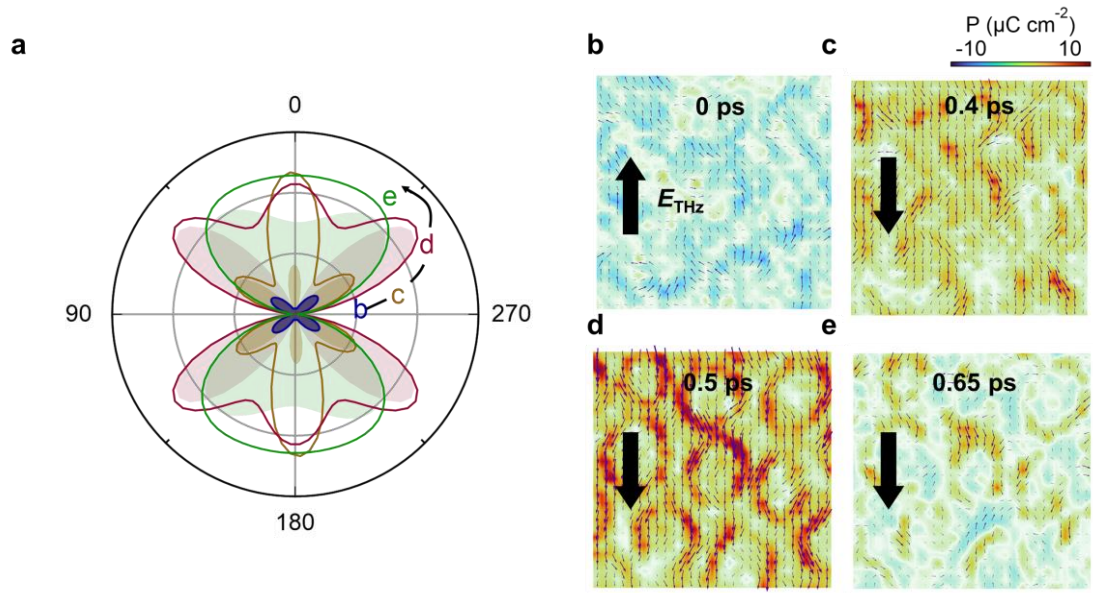

**Fig. S5. Time-domain evolution of the polar skyrmions related to the polar phase transition**
**a** SHG polarimetry plots measured at selected delay times, as marked in (**a**), with the $\chi^{(2)}$ contribution highlighted by shadow. **b - e** (001)-plane slices of the polarization configurations of the skyrmions obtained from dynamical phase-field simulations.

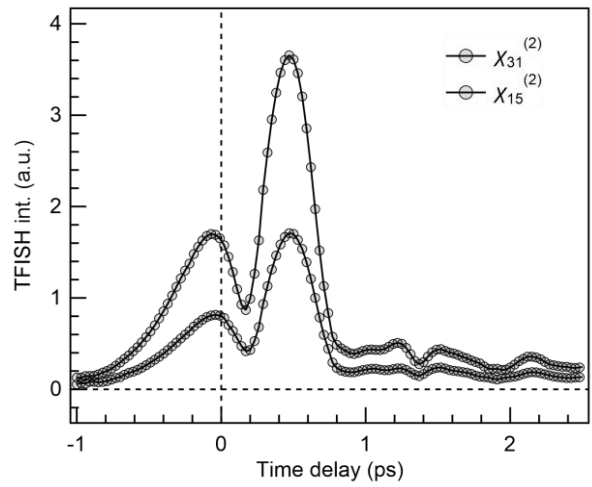

**Fig S6. Model fitting results of the $\chi^{(2)}$ contribution**

The $\chi^{(2)}$ contribution in the TFISH spectra was fitted using the equation in the main text. Similar behaviors were seen for both $\chi_{31}$ and $\chi_{15}$, showing two intensity peaks related to the rise and fall of the polar order within the skyrmions and no oscillation behavior.

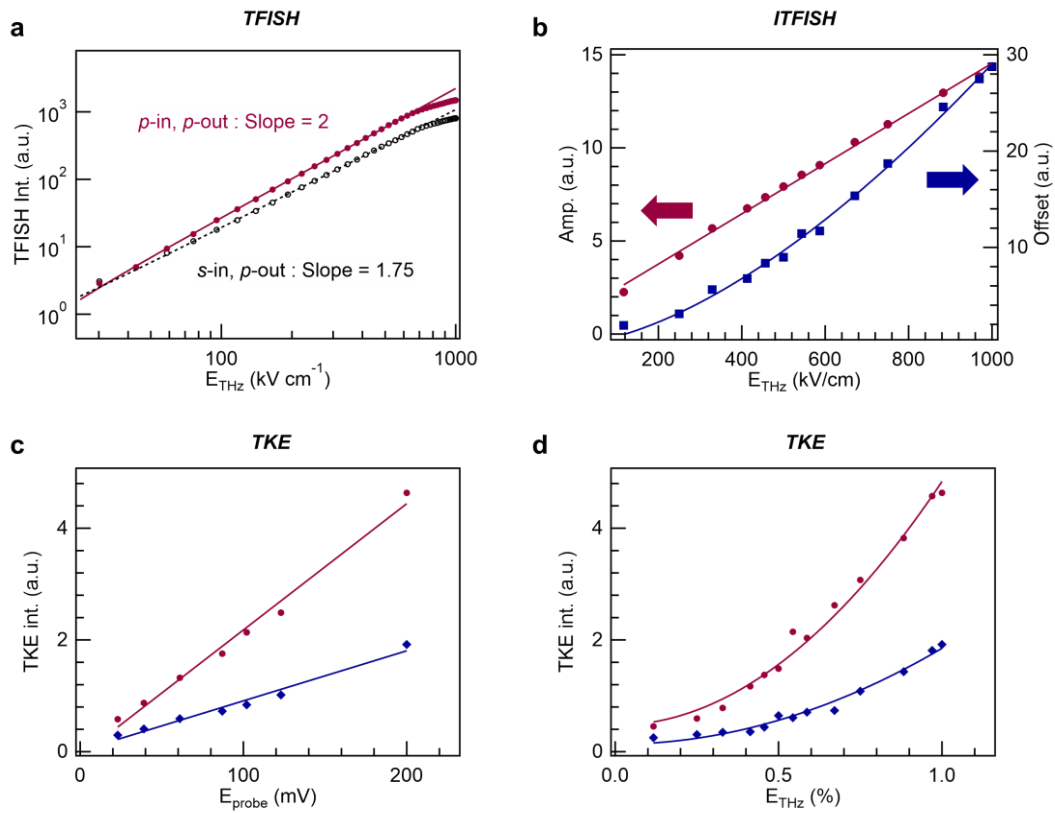

**Fig S7. The field dependence of TFISH, iTFISH and TKE**
**a** THz electric field dependence of the TFISH responses of the polar skyrmions measured at 1.7 ps along with power-function fitting results, shown in the double-logarithmic scales. **b** THz electric field dependence of the iTFISH responses. The amplitudes (red, *left*) and offsets (blue, *right*) were calculated according to the interferometric spectrums at different THz field strengths. **c** Probe beam power dependence of the TKE responses. The two sets of data were obtained individually in two measurements. **d** THz electric field dependence dependence of the TKE responses.

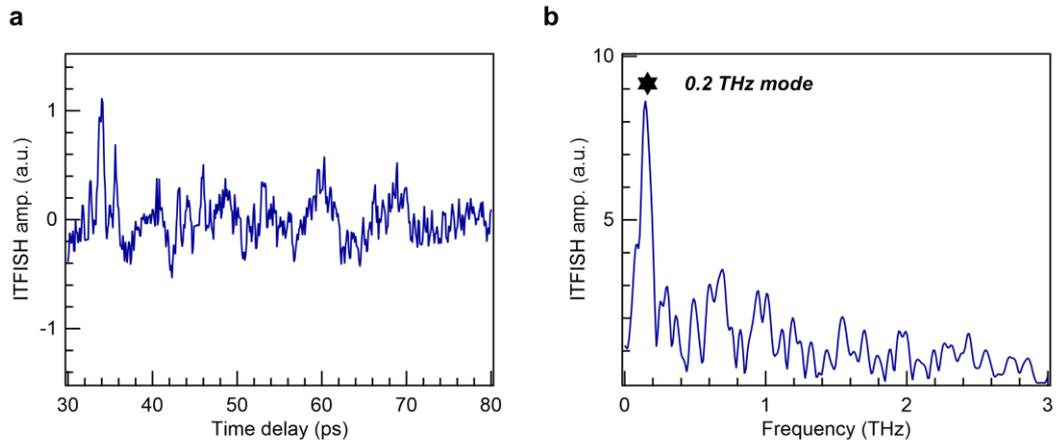

**Fig S8. Observation of the long-lasting low-frequency mode**
**a** Time-domain spectrum of the TFISH amplitude calculated from iTFISH measurements in the 30-80 ps range, showing multiple weak oscillations. **b** The Fourier-transformed spectrum of (a).

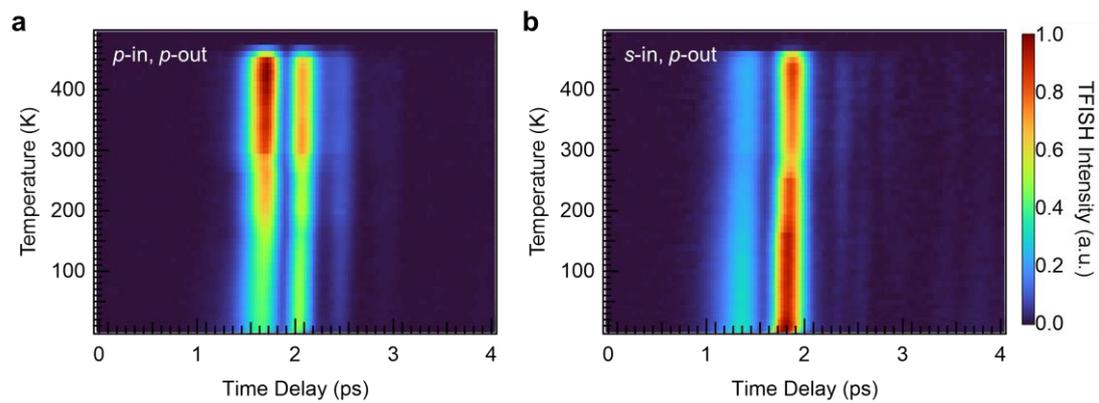

**Fig. S9. Temperature dependency of the TFISH time-domain spectra**
Temperature-dependent TFISH time domain spectroscopy under (**a**) *p*-in, *p*-out and (**b**) *s*-in, *p*-out conditions.

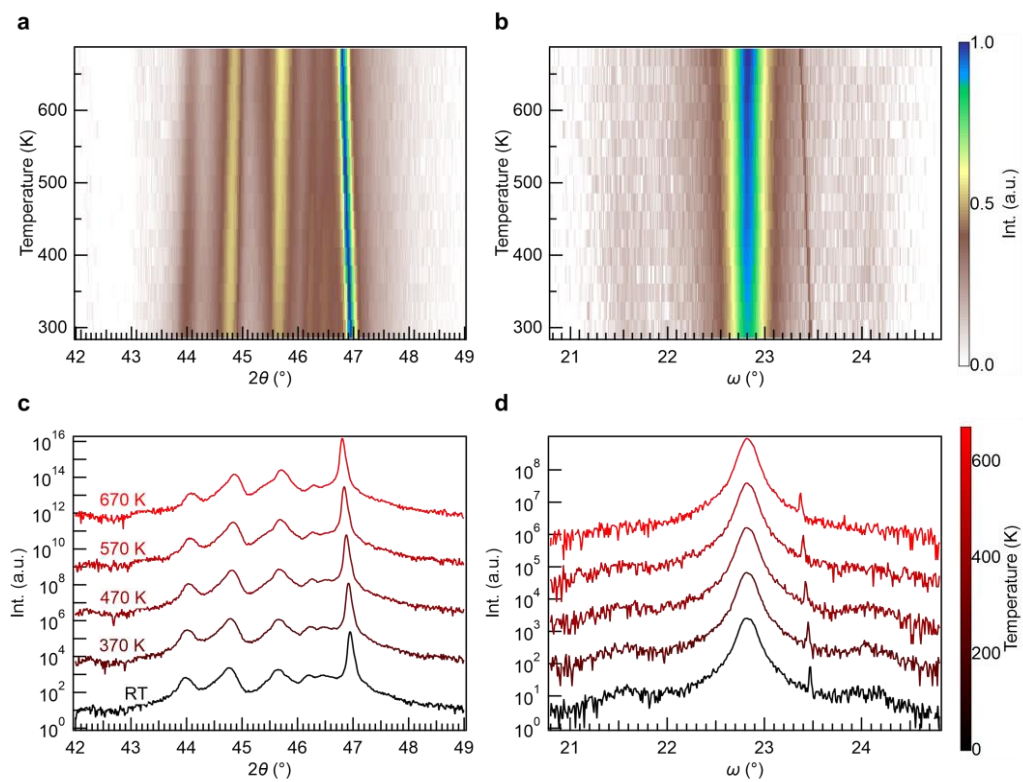

**Fig. S10. Temperature-dependent X-ray diffraction results**
Out-of-plane X-ray diffraction patterns at different temperatures measured in (**a**) $\theta$-$2\theta$ and (**b**) $\omega$-rocking modes. **c** - **d** Corresponding X-ray pattern at 5 typical temperatures.

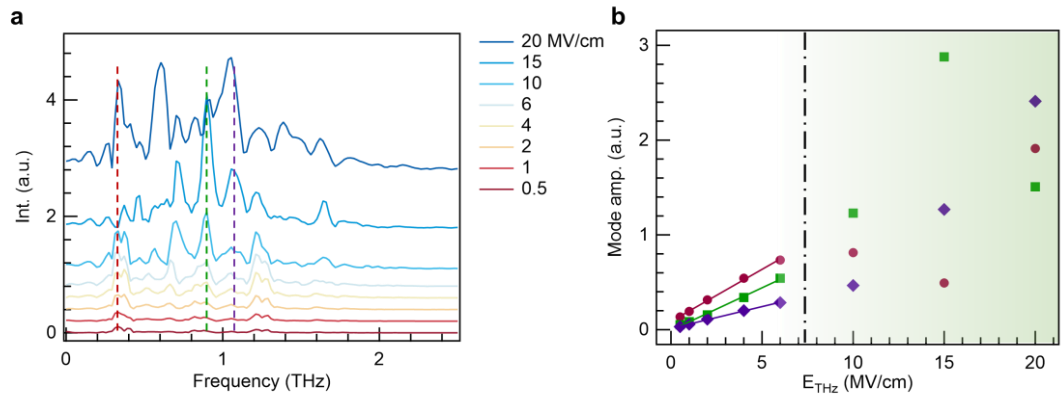

**Fig. S11. Phase-field simulation of the phonon nonlinearity**
**a** The Fourier-transformed spectra of the simulated TFISH responses at various THz field strengths. **b** The amplitude of the phonon modes as a function of THz field strength. The phonon modes at 0.35 THz, 0.9 THz and 1.05 THz are colored in red, green and violet, respectively, as marked in (**a**). The nonlinearity region starts above 6 MV/cm.

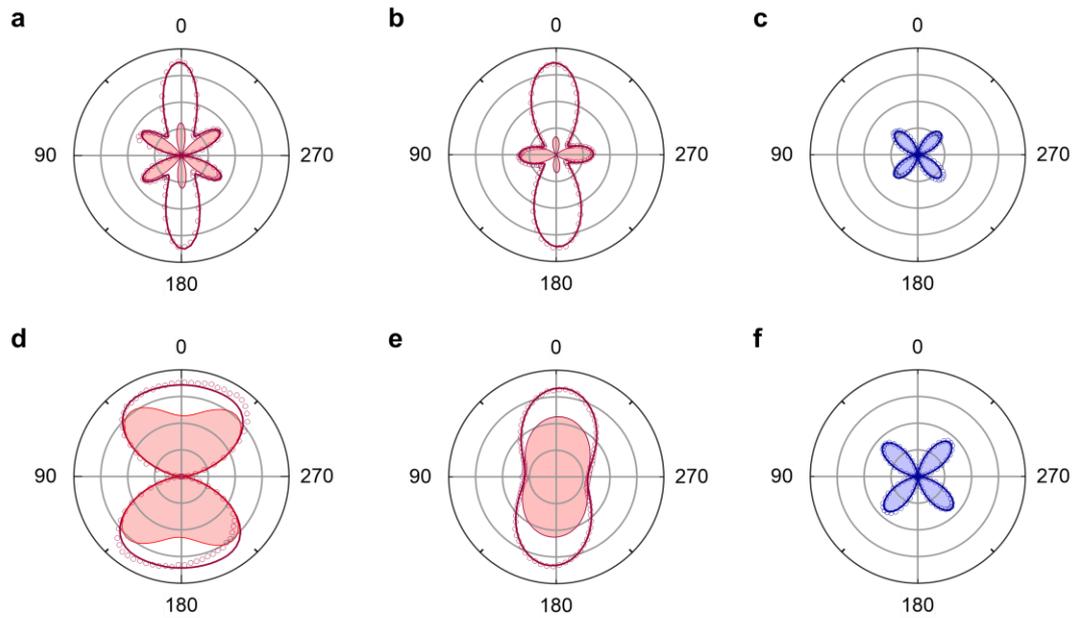

**Fig S12. SHG polarimetry results of the skyrmion at different timestamps**
SHG polarization dependence measured at **(a – c)** first TFISH peak around 1.5 ps and **(d – f)** second TFISH peak around 1.8 ps. The shadowed areas are fitted SHG contributions to the total patterns. Left panels: Results in the $\alpha=\varphi$ configuration. Middle panels: Keep $\varphi=0°$ and rotate only the incident polarization by $\alpha$. Right panels: Keep $\varphi=0°$ and rotate only the incident polarization by $\alpha$.

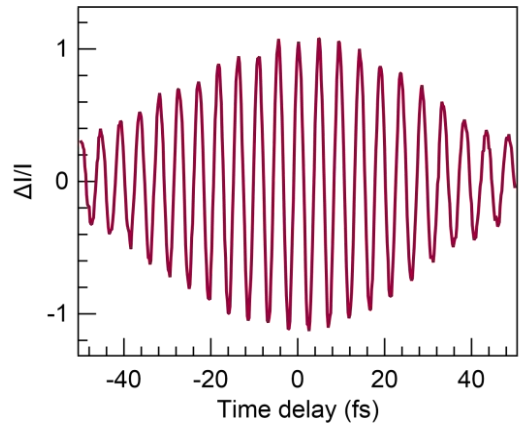

**Fig. S13. Interference pattern between TFISH responses and LO**
The iTFISH signal change as a function of the time delay between the TFISH and the LO adjusted by a pair of fused-silica wedges.